%% file: Main.tex
\documentclass{article}

\usepackage[authoryear,sort]{natbib}

\usepackage{arxiv}

\usepackage[utf8]{inputenc} % allow utf-8 input
\usepackage[T1]{fontenc}    % use 8-bit T1 fonts
\usepackage{url}            % simple URL typesetting
\usepackage{booktabs}       % professional-quality tables
\usepackage{amsfonts}       % blackboard math symbols
\usepackage{nicefrac}       % compact symbols for 1/2, etc.
\usepackage{microtype}      % microtypography
\usepackage{lipsum}		% Can be removed after putting your text content
\usepackage{graphicx}
\usepackage{natbib}
\usepackage{doi}
\usepackage{amsmath}
\usepackage{caption}
\usepackage{subcaption}
\usepackage{placeins}

\usepackage{color}
\usepackage{soul}

\title{Characterizing Lagrangian particle dynamics in decaying HIT using proper orthogonal decomposition}

\author{ \href{https://orcid.org/0000-0002-0106-6808}{\includegraphics[scale=0.06]{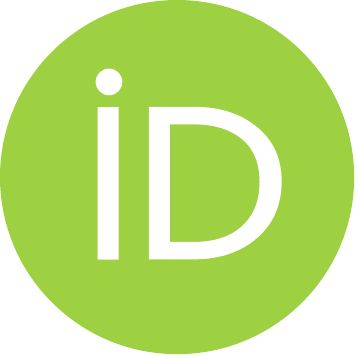}\hspace{1mm}Martin Schiødt} \\
	Technical University of Denmark\\
	Kongens Lyngby, Denmark \\
	\texttt{maschi@mek.dtu.dk} \\
	\And
	Azur Hodzic \\
	Technical University of Denmark\\
	Kongens Lyngby, Denmark \\
	\texttt{azuhod@mek.dtu.dk} \\
	\And
	Fabien Evrard \\
	Otto von Guericke University \\
	Magdeburg, Germany \\
	\texttt{fabien.evrard@ovgu.de} \\
	\And
	Berend van Wachem \\
	Otto von Guericke University \\
	Magdeburg, Germany \\
	\texttt{berend.vanwachem@ovgu.de} \\
	\And
	Clara M. Velte \\
	Technical University of Denmark \\
	Kongens Lyngby, Denmark \\
	\texttt{cmve@dtu.dk} \\
}

\renewcommand{\shorttitle}{Particle POD}

\hypersetup{
pdftitle={Characterizing Lagrangian particle dynamics in decaying HIT using proper orthogonal decomposition},
pdfsubject={q-bio.NC, q-bio.QM},
pdfauthor={Martin Schiødt},
pdfkeywords={Particle Proper Orthogonal Decomposition, Lagrangian Particles, Multiphase Flow, Reduced Order Modelling},
}

\begin{document}
\maketitle

\input{Sections/Abstract}

% keywords can be removed

\section{Introduction}
\input{Sections/Introduction}

\section{Method} \label{sec:Method}
\input{Sections/Method}

\section{Results \& discussion} \label{sec:Results}
\input{Sections/Results}

%\section{Further work} \label{sec:Discussion}
%\input{Sections/FurtherWork}

\section{Conclusion} \label{sec:Conclusion}
\input{Sections/Conclusion}

\section*{Acknowledgements}
\input{Sections/Acknowledgements}

\section*{Declaration of interest}
The authors report no conflicts of interest. 

\input{Sections/References}

\end{document}

%% file: Sections/Abstract.tex
\begin{abstract}
	The particle proper orthogonal decomposition (PPOD) is demonstrated on cases of particle flows in decaying homogeneous isotropic turbulence. Data is generated through one-way coupled simulations, where particle positions and velocities are integrated forward in time in a Lagrangian manner. The PPOD offers a direct way of extracting statistical information on the dispersed (discrete) phase of multiphase flows without any underlying assumptions. Furthermore, the method gives the possibility of modal analysis of fluid-particle interactions in multiphase flows, an example of which is provided in this work. The results demonstrate a proof of concept of the PPOD, and potential of applicability. Additionally, the results suggest that the PPOD-modes can be used for approximating particle trajectories/velocities within turbulent flows.
\end{abstract}

\iffalse
Maybe write how many modes (pct) are needed to reconstruct a certain percentage of energy.
Should not exceed 200 words.
\fi

%% file: Sections/Introduction.tex
A viscous fluid laden with particles is a dynamical system complexified by the exchange of momentum between the carrier (fluid) and dispersed (particle) phase (\cite{Schneiders2017}). Understanding these dynamics is crucial in the development of mathematical models used for turbulence modelling which in turn is important in numerous engineering applications, where the carrier phase interacts with the dispersed phase (\cite{Ferrante2003}; \cite{Burton2005}).

Since \cite{Lumley1967} introduced the \textit{classical} proper orthogonal decomposition (POD) to the fluid dynamics community, the method has served as a key statistical tool for extraction of coherent structures within turbulent fluid flows. Both within single and multiphase flows has the classical POD been applied for the extraction of dominant turbulent flow features. The past several decades have seen the birth of many modifications to the classical POD. \cite{Boree2003} introduced the \textit{extended} POD as a method of extracting flow features by not only correlating velocities, but by allowing the correlation of "any" physical quantities, such as temperature and velocity, and more recently the \textit{spectral} POD has regained traction as a method for the extraction of spatio-temporal orthogonal modes (\cite{Towne2018}). These are but a few of many POD variants, and common to all of these approaches is their applicability on the continuous (carrier) phase of fluid flows.

In recent years POD has been used in an increasing manner to study multiphase flows. \cite{Allery2005} studied the dispersion of particles within a square domain by a Galerkin projection of the Navier-Stokes equations onto POD eigenfunctions. They later extended this analysis to three dimensions (\cite{Allery2008}). In the analysis of slug flows, POD was used to extract flow patterns and transitions (\cite{Viggiano2018}) and cross-sectional phase distributions (\cite{Olbrich2021}). 

When it comes to the decomposition of the discrete (dispersed) phase of a multiphase flow, however, these methods fall short. As such, only by analyzing the carrier phase POD-modes at various flow configurations (particle volume fraction, particle-fluid density ratio, etc.) are the methods able to indirectly extract information on the dispersed phase. \cite{Pang2013} carried out such an analysis when they determined the influence of bubbles (gaseous particles) within a turbulent channel flow, by using POD on experimental data, and \cite{Higham2020} did a similar demonstration on "bubble-infected" fluidized granular flows, highlighting the utility of POD for such an analysis. 

Although the exclusive application of POD on the carrier phase can yield useful insights about both phases of multiphase flows, the remarks above still motivate the need for statistical methods, directly applicable to the dispersed phase. Recently \cite{Li2021} approached this problem by introducing \textit{Lanczos} POD (\cite{Fahl2001}) for the identification of mechanisms in powder mixing. In their application POD was used with particle position as target variable to track the displacement of a cluster of particles at sample times within a rotating paddle mixer. Equivalently we apply POD to the dispersed phase of a turbulent particle-laden flow to extract a set of temporal modes on which particle velocities may be projected. This application is termed the \textit{particle} POD (PPOD), and it is an ensemble-average based method that is applicable not only to statistically stationary but also non-stationary flows. The extracted modes are orthogonal with respect to a temporal inner product and they represent the particle kinetic energy. When applied jointly to fluid and rigid particles, the method may yield insights on fluid-particle dynamics.

The mathematical foundation of PPOD is outlined in \autoref{sec:Method} along with the data generating process used to verify the method. Results are presented and discussed in \autoref{sec:Results}. Finally, our conclusion is given in \autoref{sec:Conclusion}

%% file: Sections/Method.tex
In the following \autoref{sec:PPOD} the particle proper orthogonal decomposition is introduced along with the theoretical outline hereof. Section \ref{sec:Data} outlines the data generating process, which forms a basis of subsequent results. Finally, section \ref{sec:EnergyTrans} introduces an equation to quantify the difference between fluid and particle velocities based on modal decomposition.

\subsection{Particle POD} \label{sec:PPOD}
Let $\left\{ u^{(i)} \right\}_{i=1}^{N_e}$ be an ensemble of $N_e$ realizations. A realization, $u^{(i)}$, is defined by a collection of $N_p$ particle velocity functions,
\begin{equation}
    u^{(i)}(t) = \left\{ v_p ^{(i)} (t) \right\} _{p=1}^{N_p}\,, \quad i \in[1:N_e]\,,
\end{equation}
where $v_p ^{(i)}: T \rightarrow \mathbb{R}^3$ denotes the fluctuating (Reynolds decomposed) velocity function of particle $p$. Here $v_p ^{(i)}(t)$ is a short notation for $v_p ^{(i)}(t;z_{p,0} ^{(i)} )$, where parameter, $z_{p,0} ^{(i)} = (x_{p,0} ^{(i)}, v_{p,0} ^{(i)} )$, denotes the initial conditions of particle $p$. Specifically $x_{p,0} ^{(i)} \in \mathbb{R}^3$ is the initial position and $v_{p,0} \in \mathbb{R}^3$ is the initial velocity. The notation $v_p ^{(i)}(t) = v_p ^{(i)}(t;z_{p,0} ^{(i)})$, is chosen in order to clarify that particle velocities are functions of time only, and that particles are thus considered in a \textit{Lagrangian} frame of reference.

For a given sample point $t_k$, $k \in [0:N_t -1]$, it is noted that $u^{(i)} (t_k) \in \mathbb{R}^{3N_p}$, where the first three entries are given by $v_1^{(i)} (t_k)$, the next three by $v_2^{(i)} (t_k)$ and so on. More formally $u^{(i)} \in V$, $i\in[1:N_e]$ where
\begin{equation}
     V := \left\{ f:T \rightarrow \mathbb{R} ^{3N_p} \mid ||f|| < \infty \right\},
\end{equation}
where $T := [t_0: t_f]$ is the finite temporal domain on which we operate. $V$ is a Hilbert space with respect to the inner product $( \cdot, \cdot ) : V \times V \rightarrow \mathbb{R}$, defined as
\begin{equation}
    \left( u^{(i)}, u^{(j)} \right) = \int _T \overline{u^{(i)}(t)} \cdot u^{(j)} (t)  \mathrm{d}t\,,\quad i,j\in[1:N_e]\,, \label{eq:InnerProduct}
\end{equation}
and the norm 
\begin{equation}
    \left\| u^{(i)}\right\| = \sqrt{\left(u^{(i)},u^{(i)}\right)}.
\end{equation}
The operator $\overline{(\cdot)}$ in equation \eqref{eq:InnerProduct} refers to the complex conjugate transpose (or just transpose since the operation occurs in real space), and "$\cdot$" is the dot product. However, for ease of notation we will not write "$\cdot$" explicitly in the following. Now let $\{ \varphi _\alpha \} _{\alpha = 1}^\infty$ be a set of basis functions that spans $V$. That is, 
\begin{align}
    V = \text{span}\left\{ \varphi _\alpha \right\} _{\alpha = 1}^\infty.
\end{align}
With PPOD an orthonormal basis, $\{ \varphi _\alpha \}_{\alpha = 1}^{N_m}$, is sought, such that the expansions
\begin{align}
    u^{(i)}_{N_m} (t) = \sum \limits _{\alpha = 1} ^{N_m} c_\alpha ^{(i)} \varphi _\alpha (t),
\end{align}
of ensemble members, $u^{(i)}$, are more optimally expanded in terms of the kinetic energy by $\{ \varphi _\alpha \}_{\alpha = 1}^{N_m}$ than by any other bases. The optimality criterion can be formulated in terms of the minimization problem
\begin{align}
    \min_{ \left\{ \varphi _\alpha \in V \right\}_{\alpha=1} ^{N_m} } \left\langle \left\{ \left|\left| u^{(i)} - u^{(i)}_{N_m} \right|\right|^2  \right\}_{i=1}^{N_e} \right\rangle. \label{eq:MinimizationProblem}
\end{align}
This formulation is equivalent to the one seen in \cite{Lumley1967} and \cite{HolmesLumleyBerkooz}. Here $\langle \left\{ \cdot \right\}_{i=1}^{N_e} \rangle$ denotes the ensemble average of the $N_e$ ensembles members, so what is minimized is in other words the ensemble average of the squared norm of the difference between $u^{(i)}$ and its truncated expansion $u^{(i)}_{N_m}$. Following the same procedure as in \cite{HolmesLumleyBerkooz} we find that the optimal basis $\{ \varphi _\alpha \}_{\alpha = 1}^{N_m}$, is obtained by solving the eigenvalue problem
\begin{align}
    \mathcal{R}\varphi _\alpha (t) = \int \limits_T K(t,t') \varphi _\alpha (t') \mathrm{d}t' = \lambda _\alpha \varphi _\alpha (t)\,, \quad \alpha \in [1:N_m], \label{eq:TheEigenValueProblem} 
\end{align}
where
\begin{align}
    K(t,t') = \left\langle \left\{ u^{(i)}(t) \overline{u^{(i)}(t')} \right\}_{i=1}^{N_e} \right\rangle, \quad t,t' \in T\,, \label{eq:TheKernel}
\end{align}
and 
\begin{align}
    \lambda _1 \geq \lambda _2 \geq \cdots \geq \lambda _{N_m} \geq 0\,. \label{eq:EigenValueSequence}
\end{align}
Equation \eqref{eq:TheEigenValueProblem} is a Fredholm integral equation of the second kind, and it shows $\mathcal{R}: V \rightarrow V$ as a Fredholm integral operator, and $K(t,t')$ as a Fredholm integral equation kernel. Solving \eqref{eq:TheEigenValueProblem} is a matter of finding the eigenvalues and eigenfunctions for the symmetric kernel $K(t,t')$.

\subsubsection{PPOD in practice}
In the numerical implementation, particle velocities are only available at discrete sample times, $t_k$, $k\in [0:N_t-1]$. This means that instead of extracting continuous basis functions, it is only possible to extract a set of basis vectors from the data set given by the ensemble of discrete realizations. In this framework a realization, $u^{(i)} \in \mathbb{R}^{3 N_p N_t}$, may be ordered such that the first $3 N_p$ entries are given by $u^{(i)} (t_0)$, the next $3N_p$ entries by $u^{(i)} (t_1)$ and so on. The discrete form of the eigenvalue problem \eqref{eq:TheEigenValueProblem} is given by
\begin{align}
    R_d \varphi _\alpha = K_d \varphi _\alpha = \lambda _\alpha \varphi _\alpha\,, \quad \alpha \in [1:N_m]\,,
\end{align}
with 
\begin{align}
    K_d = \left\langle \left\{ u^{(i)} \overline{u^{(i)}}\right\}_{i=1}^{N_e} \right\rangle \in \mathbb{R}^{3N_p N_t \times 3N_p N_t}. \label{eq:DiscreteKernel}
\end{align}
An eigendecomposition of $K_d$ yields the set of basis vectors $\left\{ \varphi _\alpha \right\} _{\alpha = 1}^{N_m}$, where each basis vector, $\varphi _\alpha$, $\alpha \in [1:N_m]$, contains a modal value for all $N_p$ particles at all $N_t$ sample times in each spatial direction $x$, $y$ and $z$. As $N_p$ and $N_t$ increases, the eigendecomposition of $K_d$ may be computed efficiently using the \textit{method of snapshots} (\cite{Sirovich1987}). However, contrary to the original work where a snapshot is given by the fluid velocity at a sample time, the PPOD snapshots are constituted by the realizations, $u^{(i)}$, $i \in [1:N_e]$.

\subsection{Simulation} \label{sec:Data}
The particle velocity data used in the current application of PPOD is based on a one-way coupled simulation where fluid forces affect particles suspended in the fluid but not vice-versa. The simulation follows the Euler-Lagrange point-particle approach (\cite{Elghobashi1992}, \cite{Elghobashi1994}) which here is broken down into two stages
\begin{enumerate}
    \item Simulation of the incompressible fluid flow on a fixed Eulerian mesh, using a second-order finite-volume flow solver (\cite{Denner2020}).
    \item Simulation of the particles' motion by integrating their governing equations forward in time. To this end, the fluid velocity is interpolated linearly, both in space and time, from the discrete Eulerian velocity field.
\end{enumerate}
The fluid flow simulation uses the setup of \cite{Mallouppas2017}, who investigated various cases of forced homogeneous isotropic turbulence (HIT). Similarly, the flow field in the current case is simulated on a periodic domain discretized into $128^3$ computational cells, with spatial dimensions $l_x = l_y = l_z = 0.128$m, fluid viscosity $\mu _f = 1.7199 \times 10 ^{-5}$Pa. s. and fluid density $\rho _f = 1.17$kg m$^{-3}$. The time between each evaluation of fluid velocity is $\delta _t = 5 \times 10^{-4}$s. However, unlike \cite{Mallouppas2017} the current case is one of decaying HIT, thus the characteristic length- and time scales are non-stationary. In Table \ref{tab:SimulationParameters} the fluid specific numerical parameters of the simulation are listed, along with the Kolmogorov length- and time scale ($\eta $ and $\tau _\eta$) computed at $t=0$.
The motion of an individual particle is computed using Newton's second law of motion
\begin{align}
    m_p \frac{\mathrm{d}V_p}{\mathrm{d}t} = \sum _i F_i
\end{align}
where $F_i$ are the forces acting on the particle and $V_p$ is the particle velocity (non-Reynolds decomposed). In the present work particles are considered as rigid and spherical with diameter, $d_p$, and density, $\rho _p$. Under these conditions, the forces acting on a particle may be approximated as in the Basset-Boussinesq-Oseen equation (\cite{Basset1888}), \cite{Boussinesq1885}, \cite{Oseen1910}).
In many types of flows, the drag force, $F_{drag}$, is the most significant fluid force contribution acting on the particle, therefore we limit the present study to drag-dominated particle dynamics. The particle equation is formulated as
\begin{align}
    \theta_p \rho_p \frac{\mathrm{d}V_p}{\mathrm{d}t} = F_{drag} = \frac{\pi d_p^2}{8} \rho _f C_D ||V_f - V_p|| (V_f - V_p), \label{eq:ParticleMotion1}
\end{align}
where $V_f$ denotes the local fluid velocity. $\theta_p$ is the particle volume and $C_D$ is the drag coefficient given by (\cite{Schiller1933})
\begin{align}
    C_D = \begin{cases}
    \frac{24}{\text{Re}_p} \left( 1 + 0.15 \text{Re}_p ^{0.687} \right) \quad & \text{if } \text{Re}_p \leq 1000 \\
    0.44 \quad & \text{if } \text{Re}_p > 1000
    \end{cases}.
\end{align}
Here $\text{Re}_p$ is the particle Reynolds number computed by
\begin{align}
    \text{Re}_p = \frac{d_p \rho _f ||V_f - V_p||}{\mu _f}. \label{eq:ParticleMotion3}
\end{align}
We note that our choice of modelling using only the drag force is a crude one and that other force contributions may have to be accounted for, depending on the properties of the carrier and dispersed phase. However, this choice is advantageous both for its straightforward implementation and ability to produce reasonably realistic and physical results. These simulations may then be used to test the PPOD method and achieve preliminary results. Equations \eqref{eq:ParticleMotion1}-\eqref{eq:ParticleMotion3} are implemented along with the Runge-Kutta-based \texttt{ode23} solver in MATLAB, as to integrate particles forward in time within the periodic domain of our fluid.

\begin{table}[t]
\caption{Simulation parameters and Kolmogorov length- and time scale computed at $t=0$. The particle density $\rho _p$ is left blank, since this a parameter that will altered between case-studies.\label{tab:SimulationParameters}}
\centering
\vspace{0.1cm}
\begin{tabular}{l l l l| l l l }
 \hline\hline
  & & Fluid & &  & Particle & \\[0.5ex]
 \hline
 $\mu _f$ [Pa. s.] & $\rho _f$ [kg/m$^{3}$] & $\eta$ [m] & $\tau _\eta$ [s] & $V_p$ [m$^3$] & $\rho _p$ [kg/m$^{3}$] & $d_p$ [m] \\
 $1.7199 \times 10 ^{-5}$ & $1.17$ & $3.8441 \times 10^{-4}$ & $8.7 \times 10^{-3}$ & $1.6175 \times 10 ^{-13}$ & - & $6.76 \times 10^{-5}$ \\
\hline 
\end{tabular}
\end{table}

\subsection{Fluid and particle interactions} \label{sec:EnergyTrans}
Consider two particle velocity ensembles, $\{ u_f ^{(j)} \}_{j=1}^{N_e}$, and $\{ u_p ^{(i)} \}_{i=1}^{N_e}$. Here $u_f ^{(j)}$ denotes the $j$'th realization where the particles in question are \textit{fluid particles}. As in \cite{Yeung1988}, we define a fluid particle by the property that it assumes the local instantaneous velocity of the fluid continuum at all times. Conversely, $u_p ^{(i)}$ is the $i$'th realization of a set of rigid particles accelerated according to \eqref{eq:ParticleMotion1}-\eqref{eq:ParticleMotion3}. Using PPOD on each ensemble extracts two basis sets: $\{ \psi _\alpha \}_{\alpha = 1}^{N_m}$ and $\{ \varphi _\beta \}_{\beta = 1}^{N_m}$. Here $\psi _\alpha$, $\alpha \in [1:N_m]$, are the modes that are energy-optimal with respect to spanning fluid particle velocities and $\varphi _\beta$, $\beta \in [1:N_m]$ are the optimal modes for the rigid particles. With these bases we may write
\begin{align}
    u_f ^{(j)}(t) &= \sum \limits_{\alpha = 1}^{N_m} (u_f ^{(j)} , \psi _\alpha) \psi _\alpha (t) = c^{(j),\alpha} \psi _\alpha (t)\,, \quad j \in [1:N_e]\,,\\
    u_p ^{(i)}(t) &= \sum \limits_{\beta = 1}^{N_m} (u_p ^{(i)}, \varphi _\beta) \varphi _\beta (t) = d^{(i),\beta} \varphi _\beta (t)\,, \quad i \in [1:N_e]\,.
\end{align}
where repeated indices in a term imply the Einstein summation convention. The average of the residual norm squared between fluid- and rigid particle realizations reduces to
\begin{align}
    \left\langle \left\{ ||u^{(i)}_f-u^{(i)}_p||^2 \right\}_{i=1}^{N_e} \right\rangle = \sum \limits_{\alpha = 1}^{N_m} \lambda _\alpha + \sum \limits_{\beta = 1}^{N_m} \sigma _\beta - 2 \left\langle \left\{ c^{(i),\alpha} d^{(i),\beta}  \right\} _{i=1}^{N_e} \right\rangle (\psi _\alpha, \varphi _\beta )\,, \label{eq:NormedDiff}
\end{align}
where $\lambda _\alpha$ and $\sigma _\beta$, $\alpha, \beta \in [1:N_m]$, are the eigenvalues associated with $\psi _\alpha$ and $\varphi _\beta$ respectively. \\
In the following section the inner product (third term) of equation \eqref{eq:NormedDiff} is analyzed, where the initial position of the fluid particles in $u_f ^{(j)}$ are the same as those of the rigid particles in $u_p ^{(i)}$ for $i = j$. If the trajectories (and thereby velocities) in these two types of realizations are the same, then the average normed residual is zero, meaning the bases $\{ \psi _\alpha \}_{\alpha=1}^{N_e}$ and $\{ \varphi _\beta \}_{\beta=1}^{N_e}$ are parallel. Varying for instance the density of the rigid particles, may yield information about the flow. These points will be addressed in \autoref{sec:Results}.

Consider now the case where $v^{(i)}_p(t)$ denotes not only the fluctuating part, but the full velocity of a particle. Then all of the derivations above still hold, the only difference being that the first PPOD mode would often constitute the ensemble averaged velocity, as this is typically the most energetic mode. Furthermore, let $u_f^{(i)}$ be interpreted as the fluid velocity local to the rigid particle. To avoid confusion we call this quantity $u_{f,loc}^{(i)}$. Under this framework the particle equations of motion \eqref{eq:ParticleMotion1}--\eqref{eq:ParticleMotion3} can be written in modal form. For instance, the normed difference between the local fluid velocity and the rigid particle velocity is given as
\begin{align}
    \left| \left| u^{(i)} _{f,loc} - u^{(i)} _{p} \right| \right| = \sqrt{ \left( u^{(i)} _{f,loc}, u^{(i)} _{f,loc} \right) + \left( u^{(i)} _{p}, u^{(i)} _{p} \right) - 2 \left( u^{(i)} _{f,loc}, u^{(i)} _{p} \right) },
\end{align}
which in modal form is given by
\begin{align}
    \left| \left| u^{(i)} _{f,loc} - u^{(i)} _{p} \right| \right| = \sqrt{ \sum \limits_{\alpha=1}^{N_m} \left|c ^{(i)}_{\alpha,loc} \right|^2 + \sum \limits_{\beta=1}^{N_m} \left|d ^{(i)}_\beta \right|^2 - 2 c ^{(i),\alpha} d ^{(i),\beta} (\psi _\alpha , \varphi _\beta ) }. \label{eq:DragForceFactor}
\end{align}
When $N_p = 1$ this equation can be inserted directly into \eqref{eq:ParticleMotion1}--\eqref{eq:ParticleMotion3}, and when the modes $\{\psi_\alpha\}_{\alpha=1}^{N_m}$ and $\{\varphi_\beta\}_{\beta=1}^{N_m}$ are known, this gives a concrete way of statistically approximating the (drag) force that the rigid particle experiences as it travels through the fluid. 

\begin{figure}
\centering
\begin{subfigure}{.5\textwidth}
  \centering
  \includegraphics[scale=0.49]{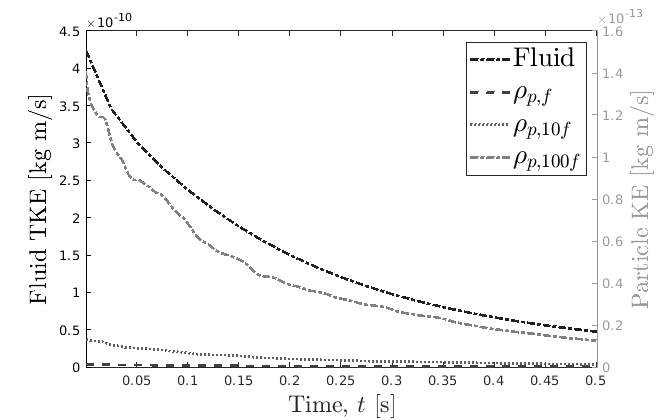}
  \caption{Kinetic energy}
  \label{fig:FluidKineticEnergy}
\end{subfigure}%
\begin{subfigure}{.5\textwidth}
  \centering
  \includegraphics[scale=0.48]{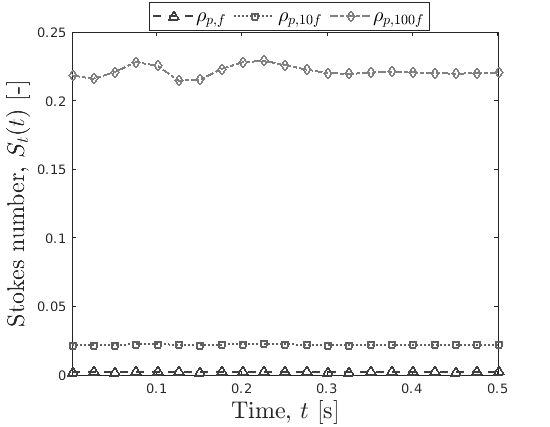}
  \caption{Stokes number}
  \label{fig:StokesNumber}
\end{subfigure}
\caption{(a) The TKE of the fluid and the ensemble averaged kinetic energy of the single particle over time (three different densities $\rho_{p,f}$ and $\rho_{p,10f}$ and $\rho_{p,100f}$). (b) The Stokes number, $S_t(t)$, in the three different density cases as it evolves over time.}
\label{fig:FluidAndStokes}
\end{figure}
\FloatBarrier

%% file: Sections/Results.tex
This section consists of the analysis of two cases - the single particle realization (SPR) case and the multiple particle realization (MPR) case. Only $N_p = 1$ particle is integrated forward in time in the SPR case within each realization, whereas $N_p = 20$ particles are integrated forward in time in the MPR case. The analysis of these two cases is carried out in sections \ref{sec:SingleParticle} and \ref{sec:MultipleParticle}, respectively. 

In both cases the particles are initiated with the same velocity as the local fluid velocity, and $N_e = 160$ realizations are simulated consisting of $N_t = 1001$ sample times. This confines the temporal domain to $T := [t_0, t_f] = [0\text{s}, 0.5\text{s}]$, motivated by the fact that the turbulence kinetic energy (TKE) is relatively small at $0.5\text{s}$ (see Figure \ref{fig:FluidKineticEnergy}). The SPR and the MPR cases can be split into the following three substudies:
\begin{itemize}
    \item Study 1: particles have density $\rho_p = \rho_{p,f} = \rho_f$ (neutrally buoyant)
    \item Study 2: particles have density $\rho_p = \rho_{p,10f} = 10\rho_f$
    \item Study 3: particles have density $\rho_p = \rho_{p,100f} = 100\rho_f$
\end{itemize}
This means, that both the SPR and MPR cases are analyzed with varying particle densities.

\subsection{PPOD - Single particle} \label{sec:SingleParticle}
Statistically, the kinetic energy of the particles is decaying over time due to the decaying TKE of the flow field. This tendency is depicted in Figure \ref{fig:FluidKineticEnergy} for studies 1--3, where the ensemble averaged particle kinetic energy of the SPR case is shown. The difference in energy between the three particle curves is due to the density difference. In Figure \ref{fig:StokesNumber} the Stokes number, $S_t$, as a function of time is depicted, which is defined as, \cite{Mallouppas2017}
\begin{equation}
    S_t(t) = \frac{\tau _p (t)}{\tau _\eta}\,,
\end{equation}
where
\begin{equation}
    \tau _p (t) = \frac{\rho_p d_p ^2}{18 \mu _f} \frac{C_D (t) Re_p (t)}{24}\,,
\end{equation}
is the particle response time, and $\tau _\eta$ is a characteristic time scale of the fluid - in this case chosen to be the Kolmogorov time scale at $t_0$ ($\tau _\eta = 8.7 \times 10^{-3}$).

\begin{figure}
    \centering
    \includegraphics[scale=0.7]{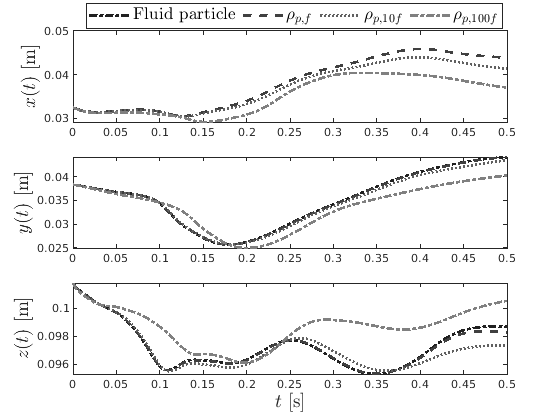}
    \caption{Particle positions as a function of time sampled from our ensemble of realizations. The plot shows that rigid particles with fluid density follow a trajectory almost identical to the fluid particle, whereas trajectories of denser particles deviate more from the fluid streamline.}
    \label{fig:SampleParticleTrajectory}
\end{figure}

Figure \ref{fig:SampleParticleTrajectory} shows the different trajectories traced out by a fluid particle and a rigid particle of different densities. Particles with fluid density, $\rho_{p,f}$, have trajectories that are almost identical to those of fluid particles, whereas trajectories of denser particles ($\rho_{p,10f}$ and $\rho_{p,100f}$) are seen to deviate more from the fluid streamline. This is because an increase in particle-fluid density ratio $\rho_p / \rho_f$ infers a decrease in the effect of the local fluid motion on the particle (\cite{Shen2021}). However, the figure still illustrates that at low sample times ($t < 0.05$s) the trajectories of the rigid particles with different densities are nearly identical, and only later do they start to deviate, with heavier particles deviating sooner than their lighter counterparts. The initial coherence between the particle trajectories may be explained by the high kinetic energy in the initial stages making the considered particle densities inconsequential in comparison.
\FloatBarrier

There are several contributions to the deviation between streamlines of neutrally buoyant rigid particles and those of the fluid flow. One contribution stems from fluid particles being considered to be infinitesimal in size, whereas the particles we are simulating are rigid and of finite size. Furthermore, the drag force with which a rigid particle is accelerated in the current work is merely an approximation of the resulting forces a solid particle is exposed to in reality -  other forces such as the "added mass"-force may be needed for a more accurate model. The reader is referred to \cite{Homann2010} for a more thorough analysis of neutrally buoyant particles in turbulent flow.

\begin{figure}
\centering
\begin{subfigure}{.5\textwidth}
  \centering
  \includegraphics[scale=0.55]{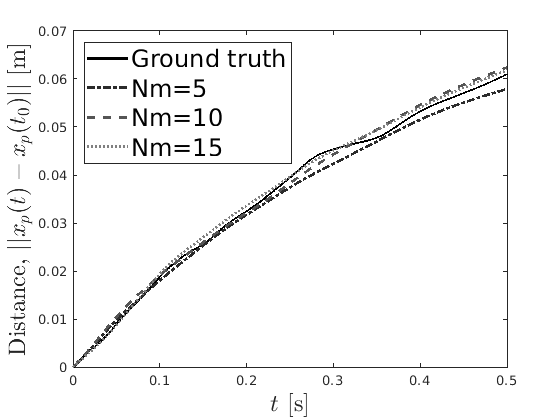}
  \caption{Displacement}
  \label{fig:AbsDistTraj}
\end{subfigure}%
\begin{subfigure}{.5\textwidth}
  \centering
  \includegraphics[scale=0.55]{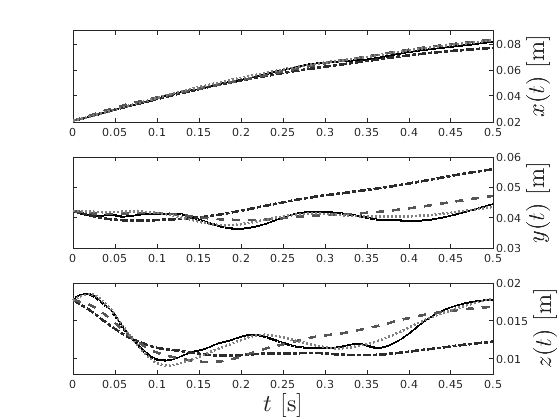}
  \caption{Position}
  \label{fig:CompTraj}
\end{subfigure}
\caption{(a) Displacement of particle over time from initial position. (b) Particle position as a function of time in $x$, $y$ and $z$ direction. Both plots display the ground truth along with cases $N_m = 5,10,15$ which are the number of PPOD modes used to reconstruct the particle trajectory.}
\label{fig:ReconstructedTrajectory}
\end{figure}

\begin{figure}
    \centering
    \includegraphics[scale=0.65]{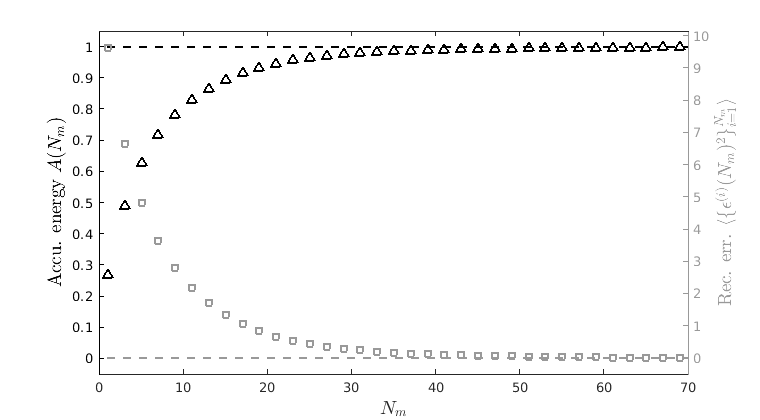}
    \caption{$A(N_m)$ (accumulated modal energy) converging towards 1 (100\%) as $N_m \rightarrow N_e$ (triangle), and average squared reconstruction error (square) converging towards zero, as the number of reconstruction modes increases.}
    \label{fig:SoloConvergence}
\end{figure}

Due to the decaying TKE of the flow field the PPOD modes reconstruct the particle velocities more rapidly in the beginning of the particle trajectory, due to the high kinetic energy associated with the particles at this stage. This is depicted in Figure \ref{fig:ReconstructedTrajectory} showing the full reconstruction of the trajectory, designated by \textit{Ground truth}, as well as for the cases $N_m = [5,10,15]$, referring to the number of PPOD modes used to reconstruct the trajectories. The trajectories are computed from
\begin{align}
    x_p (t) = x_{p}(t_0) + \int \limits_{t_0}^{t} v_p(\tau) \mathrm{d}\tau \approx x_{p} (t_0) + \int \limits_{t_0}^{t} \sum \limits_{\alpha=1}^{N_m} c_\alpha \varphi _{p,\alpha} (\tau) \mathrm{d}\tau\,.
\end{align}
Here $\varphi _{p,\alpha} (t)$ is the eigenvector component of $\varphi _{\alpha}$ related to particle $p$, where $\varphi _{p, \alpha} = \varphi _\alpha$ for $N_p = 1$. Figure \ref{fig:AbsDistTraj} shows the displacement (Euclidean distance) of the particle from its initial position and Figure \ref{fig:CompTraj} shows the component-wise position ($x$, $y$, $z$) over time.  The figure not only shows that particle position/velocity converges more rapidly at lower times, but also indicates that for increasing $N_m$, the reconstruction converges towards the ground truth. The convergence of the reconstruction is shown for study 1 in Figure \ref{fig:SoloConvergence} quantified by the ensemble average of the square of the reconstruction error of the particle velocities as a function of the number of reconstruction modes, $\langle \{ \epsilon ^{(i)} (N_m) ^2 \}_{i=1}^{N_e} \rangle$, where
\begin{align}
    \epsilon ^{(i)} (N_m) = \left|\left|u^{(i)} - u^{(i)}_{N_m}\right|\right|, \quad i = 1,\dots,N_e\,.
\end{align}
In the same figure the fraction quantifying the accumulated energy of the PPOD modes up until mode $N_m$
\begin{align}
    A(N_m) = \sum \limits_{\alpha=1}^{N_m} \lambda _\alpha \bigg/ \sum \limits_{\beta=1}^{N_e} \lambda _\beta\,,
\end{align}
is shown. It is clear from Figure \ref{fig:SoloConvergence} that equation \eqref{eq:EigenValueSequence} is satisfied, and that the first couple of modes account for a large portion of particle kinetic energy. For the sake of readability every second marker is suppressed in the figure. The quantity $\epsilon (N_m) ^2$ is a measure of the energy that remains to be reconstructed in the finite expansion $u^{(i)}_{N_m}$, and therefore 
\begin{align}
    \langle \{ \epsilon ^{(i)} (N_m) ^2 \}_{i=1}^{N_e} \rangle \rightarrow 0 \text{ for } N_m \rightarrow N_e,
\end{align}
at a rate similar to
\begin{align}
    A(N_m) \rightarrow 1 \text{ for } N_m \rightarrow N_e.  
\end{align}

\begin{figure}
\centering
\begin{subfigure}{.5\textwidth}
  \centering
  \includegraphics[scale=0.52]{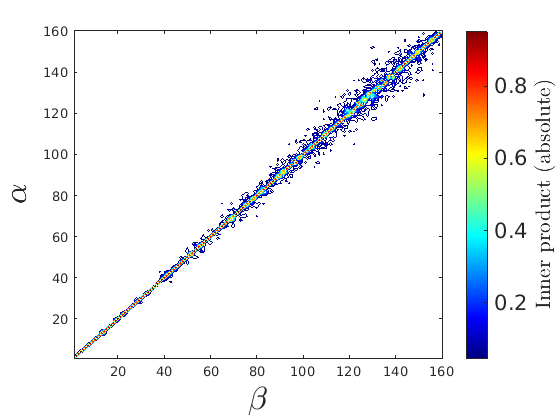}
  \caption{$\rho_{p,f}$}
  \label{fig:ContourRhopf}
\end{subfigure}%
\begin{subfigure}{.5\textwidth}
  \centering
  \includegraphics[scale=0.52]{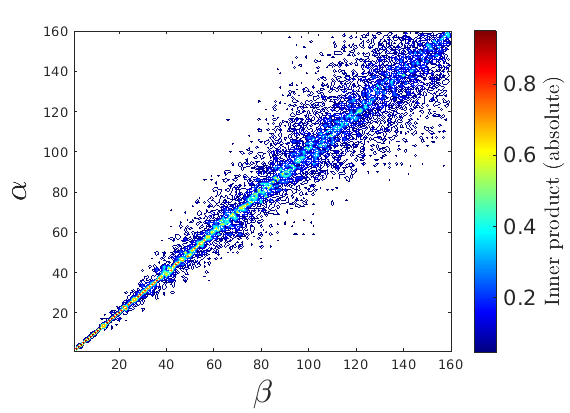}
  \caption{$\rho_{p,10f}$}
  \label{fig:ContourRhop10f}
\end{subfigure}
\begin{subfigure}{.5\textwidth}
  \centering
  \includegraphics[scale=0.52]{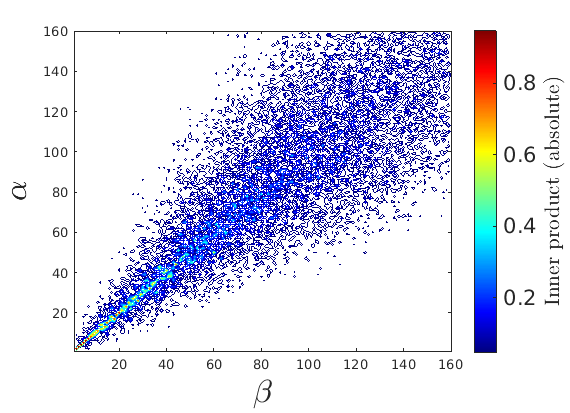}
  \caption{$\rho_{p,100f}$}
  \label{fig:ContourRhop100f}
\end{subfigure}
\caption{Contours of the absolute value of the inner product between fluid particle modes, $\{ \psi _\alpha \}_{\alpha =1}^{N_e}$, and rigid particle modes ,$\{ \varphi _\beta \}_{\beta =1}^{N_e}$. (a)--(c) shows the contour for the case of particle with density $\rho _{p,f}$--$\rho _{p,100f}$ respectively.}
\label{fig:SoloModeContour}
\end{figure}
\FloatBarrier

With regards to equation \eqref{eq:NormedDiff} the parallelity between the PPOD modes for the fluid particle and the rigid particle of different densities is evaluated. In Figure \ref{fig:ContourRhopf}-- \ref{fig:ContourRhop100f}, contours of the absolute value of the inner product of the normed fluid particle modes, $\{ \psi _\alpha \}_{\alpha =1}^{N_e}$, and the normed modes, $\{ \varphi _\beta \}_{\beta =1}^{N_e}$, of the rigid particle with varying density, is shown. Given that particles with fluid density ($\rho_{p,f}$) had trajectories almost identical to fluid particles (Figure \ref{fig:SampleParticleTrajectory}), we see that the modes of these two types of particles are predominantly parallel. This is shown in Figure \ref{fig:ContourRhopf}, where the contour of the inner product attains maximal values along the diagonal, and is decreasing rapidly in the off-diagonal region. This means that modes $\psi _\alpha$ and $\varphi _\beta$ are parallel (or at least close to) for $\alpha = \beta$. For particle densities $\rho_{p,10f}$ and $\rho_{p,100f}$ (Figure \ref{fig:ContourRhop10f}--\ref{fig:ContourRhop100f}), the first few particle modes exhibit reasonable parallelity with corresponding fluid modes, whereafter this tendency diminishes with increasing mode numbers. It is worth noting that the parallelity between the two sets of modes, generally seems highest between modes that have similar mode numbers and thereby represent similar kinetic energy levels. We can designate this feature of the dynamical system by \textit{local interactions}, where the term \textit{local} refers to the similarity of mode numbers exhibiting parallelity. Given the hypothesis that the interactions between fluid and solid particles is dominated by local interactions, these results may be used to analyze the spectral range of the interaction as a function of spectral energy. However, since this analysis is applied to a one-way coupled simulation, the physical insight of particle dynamics is limited by this fact. It does, however, demonstrate the ability and potential of the PPOD to extract insights from more complex systems, such as two- or four-way coupled systems.
\begin{figure}
\centering
\begin{subfigure}{.5\textwidth}
  \centering
  \includegraphics[scale=0.55]{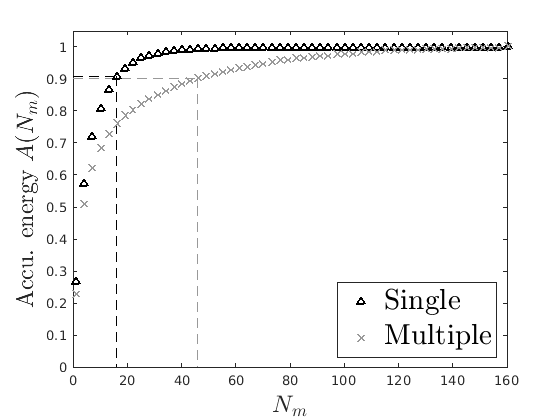}
  \caption{Single vs. Multiple}
  \label{fig:EigenvalueSoloMultiComparison}
\end{subfigure}%
\begin{subfigure}{.5\textwidth}
  \centering
  \includegraphics[scale=0.55]{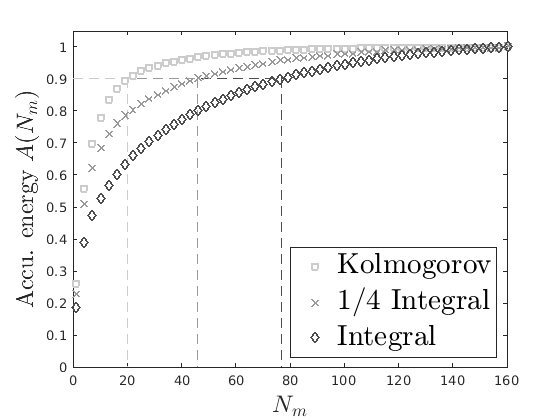}
  \caption{Varying initial separations}
  \label{fig:EigenvalueSeparationComparison}
\end{subfigure}
\caption{Convergence of accumulated modal energy, $A(N_m)$. (a) Compares how many modes are needed to account for 90\% of particle kinetic energy in the single particle and multiple particles realization cases. (b) Compares how many modes are needed to account for 90\% of particle kinetic energy when the initial separation (position) of the particles vary.}
\label{fig:EigenvalueComparison}
\end{figure}

\subsection{PPOD - Multiple particles} \label{sec:MultipleParticle}
In two-way coupled systems clustering of particles in low vorticity regions is often observed (\cite{Squires1990}, \cite{Balachandar2010}). We may expect particles initiated in close proximity to each other in such a system to follow similar streamlines. This motivates the following PPOD analysis of multiple particle realizations where particles are initiated within a confined region of the fluid flow.

In Figure \ref{fig:EigenvalueSoloMultiComparison} the number of modes required to account for 90\% of particle kinetic energy is shown for study 1 in the SPR and MPR cases. In the SPR case 10\% of the available modes cover 90\% of the particle kinetic energy, whereas $\sim$28\% is required in the MPR case. Thus it is clear that there is a faster convergence for SPR. Since the MPR PPOD incorporates velocity correlations between all particles over all times, it results in a more complex kernel, which explains the slower convergence rate of the eigenvalues than for the SPR case.

In \cite{Mehrabadi2018} the interaction of neighboring particles in decaying HIT is considered. Among the factors they consider is the relative distance between neighboring particles. We here note that the initial separation of the particles in the MPR case was chosen somewhat arbitrarily. That is, the PPOD modes extracted from the MPR case relied on realizations where particles had an initial separation of at most $\sim 1/4$ of the integral length scale at $t_0$ ($\tau _I = 8.3 \times 10^{-3}$m). If the particles have a lower initial separation then the correlation between their velocities will be higher, because smaller turbulent structures are required to separate the particles as they move through the fluid. Conversely, if the initial separation of the particles is larger then the correlation between the particle velocities becomes smaller as the particles are affected by a broader range of turbulent structures. In Figure \ref{fig:EigenvalueSeparationComparison} an indication of this behaviour is seen. The figure illustrates the number of modes needed to account for $90\%$ of particle kinetic energy when the realizations have different initial separations, where the separation range is from the Kolmogorov length scale to the integral length scale. Generally, a more rapid convergence is seen for small separations than for large ones induced by a higher correlation between particle velocities in the "Kolmgorov"-case than in the other two cases.

\begin{figure}
\centering
\begin{subfigure}{.5\textwidth}
  \centering
  \includegraphics[scale=0.55]{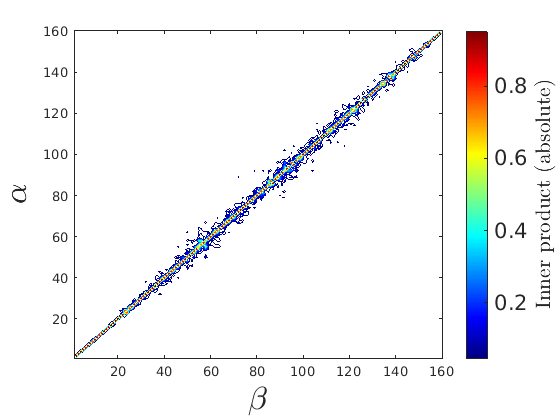}
  \caption{$\rho_{p,f}$}
  \label{fig:MultiContourRhopf}
\end{subfigure}%
\begin{subfigure}{.5\textwidth}
  \centering
  \includegraphics[scale=0.55]{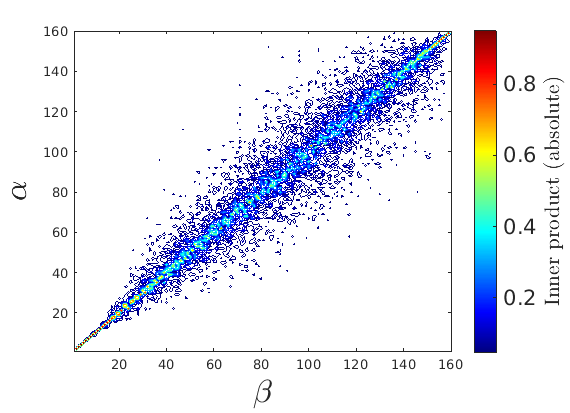}
  \caption{$\rho_{p,10f}$}
  \label{fig:MultiContourRhop10f}
\end{subfigure}
\caption{Contours of the absolute value of the inner product between fluid particle modes, $\{ \psi _\alpha \}_{\alpha =1}^{N_e}$, and rigid particle modes ,$\{ \varphi _\beta \}_{\beta =1}^{N_e}$ in the multiple particle realization case. (a) shows the contour for the case of particle densities $\rho _{p,f}$ and (b) shows the contours for the case with particle densities $\rho _{p,10f}$.}
\label{fig:MultiModeContour}
\end{figure}

The parallelity of fluid particle PPOD modes and rigid particle PPOD modes are quantified in Figure \ref{fig:MultiModeContour}. Here it is seen that fluid particle modes and rigid particle modes exhibit high parallelity when the rigid particles have density $\rho_{p,f}$. When the density increases (Figure \ref{fig:MultiContourRhop10f}) the parallelity decreases for higher mode numbers. Note here that the tendency for the MPR case (where initial separation is 1/4 integral) is not the same as for the SPR case. For instance, there seems to be an increase in parallelity in the MPR case with the fluid particles for very high mode numbers (close to 160), whereas for the SPR case the parallelity steadily decreased for still higher mode numbers. This shows that single particle and multiple particle realization PPOD-analysis may yield different insights, when applied to physical problems. 

%% file: Sections/Conclusion.tex
Particle POD is introduced as an extension of the classical POD and as a method for the extraction of temporal modes from turbulent flow fields. The method offers a way of extracting statistical information directly from the discrete phase of multiphase flows. PPOD is demonstrated on one-way coupled simulations of particles in a decaying homogeneous isotropic turbulent flow. The results show promise in analyzing fluid-particle interactions in turbulent flows using modal analysis. The strength of PPOD is that no underlying assumptions about the carrier or dispersed phase is needed, and that the method can be applied directly to decompose particle velocities in general particle-laden flows. Finally, the results suggest that the PPOD is useful for approximating particle velocities in turbulent flows.

%% file: Sections/Acknowledgements.tex
AH and CMV acknowledge financial support from the European Research council: This project has received funding from the European Research Council (ERC) under the European Unions Horizon 2020 research and innovation program (grant agreement No 803419). 

MS acknowledges financial support from the Poul Due Jensen Foundation: Financial support from the Poul Due Jensen Foundation (Grundfos Foundation) for this research is gratefully acknowledged.

%% file: Sections/References.tex
\bibliographystyle{apalike} %unsrtnat
\bibliography{references}  %%% Uncomment this line and comment out the ``thebibliography'' section below to use the external .bib file (using bibtex) .

%% file: Main.bbl
\begin{thebibliography}{}

\bibitem[Allery et~al., 2005]{Allery2005}
Allery, C., Beghein, C., and Hamdouni, A. (2005).
\newblock Applying proper orthogonal decomposition to the computation of
  particle dispersion in a two-dimensional ventilated cavity.
\newblock {\em Communications in Nonlinear Science and Numerical Simulation},
  10:907--920.

\bibitem[Allery et~al., 2008]{Allery2008}
Allery, C., Beghein, C., and Hamdouni, A. (2008).
\newblock On investigation of particle dispersion by a pod approach.
\newblock {\em International Applied Mechanics}, 44:110--119.

\bibitem[Balachandar and Eaton, 2010]{Balachandar2010}
Balachandar, S. and Eaton, J. (2010).
\newblock Turbulent dispersed multiphase flow.
\newblock {\em Annual Review of Fluid Mechanics}, 42:111--133.

\bibitem[Basset, 1888]{Basset1888}
Basset, A. (1888).
\newblock {\em A Treatise on Hydrodynamics with Numerous Examples}, volume~2.
\newblock Deighton, Bell and Company.

\bibitem[Borée, 2003]{Boree2003}
Borée, J. (2003).
\newblock Extended proper orthogonal decomposition: a tool to analyse
  correlated events in turbulent flows.
\newblock {\em Experiments in Fluids}, 35:188--192.

\bibitem[Boussinesq, 1885]{Boussinesq1885}
Boussinesq, J. (1885).
\newblock Sur la résistance qu’oppose un fluide indéfini au repos, sans
  pesanteur, au mouvement varié d’une sphère solide qu’il mouille sur
  toute sa surface, quand les vitesses restent bien continues et assez faibles
  pour que leurs carrés et produits soient négligeables.
\newblock {\em Comptes Rendus de l’Académie des Sciences}, 100:935--937.

\bibitem[Burton and Eaton, 2005]{Burton2005}
Burton, T. and Eaton, J. (2005).
\newblock Fully resolved simulations of particle-turbulence interaction.
\newblock {\em Journal of Fluid Mechanics}, 545:67--111.

\bibitem[Denner et~al., 2020]{Denner2020}
Denner, F., Evrard, F., and van Wachem, B. (2020).
\newblock Conservative finite-volume framework and pressure-based algorithm for
  flows of incompressible, ideal-gas and real-gas fluids at all speeds.
\newblock {\em Journal of Computational Physics}, 409:109348.

\bibitem[Elghobashi, 1994]{Elghobashi1994}
Elghobashi, S. (1994).
\newblock On predicting particle-laden turbulent flows.
\newblock {\em Applied Scientific Research}, 52:309--329.

\bibitem[Elghobashi and Truesdell, 1992]{Elghobashi1992}
Elghobashi, S. and Truesdell, G. (1992).
\newblock Direct simulation of particle dispersion in a decaying isotropic
  turbulence.
\newblock {\em Journal of Fluid Mechanics}, 242:655--700.

\bibitem[Fahl, 2001]{Fahl2001}
Fahl, M. (2001).
\newblock Computation of pod basis functions for fluid flows with lanczos
  methods.
\newblock {\em Mathematical and Computer Modelling}, 34:91--107.

\bibitem[Ferrante and Elghobashi, 2003]{Ferrante2003}
Ferrante, A. and Elghobashi, S. (2003).
\newblock On the physical mechanisms of two-way coupling in particle-laden
  isotropic turbulence.
\newblock {\em Physics of Fluids}, 15(2):315--329.

\bibitem[Higham et~al., 2020]{Higham2020}
Higham, J., Shahnam, M., and Vaidheeswaran, A. (2020).
\newblock Using a proper orthogonal decomposition to elucidate features in
  granular flows.
\newblock {\em Granular Matter}, 22:86.

\bibitem[Holmes et~al., 2012]{HolmesLumleyBerkooz}
Holmes, P., Lumley, J.~L., and Berkooz, G. (2012).
\newblock {\em Turbulence, Coherent Structures, Dynamical Systems and
  Symmetry}.
\newblock Cambridge Univ. Press, 2 edition.

\bibitem[Homann and Bec, 2010]{Homann2010}
Homann, H. and Bec, J. (2010).
\newblock Finite-size effects in the dynamics of neutrally buoyant particles in
  turbulent flow.
\newblock {\em Journal of Fluid Mechanics}, 651:81--91.

\bibitem[Li et~al., 2021]{Li2021}
Li, S., Duan, G., and Sakai, M. (2021).
\newblock Pod-based identification approach for powder mixing mechanism in
  eulerian-langrangian simulations.
\newblock {\em Advanced Powder Technology}, In press.

\bibitem[Lumley, 1967]{Lumley1967}
Lumley, J.~L. (1967).
\newblock The structure of inhomogeneous turbulent flows.
\newblock {\em Atmospheric Turbulence and Radio Wave Propagation - Proceedings
  of the International Colloquium - Moscow}, pages 166--176.

\bibitem[Mallouppas et~al., 2017]{Mallouppas2017}
Mallouppas, G., George, W., and van Wachem, B. (2017).
\newblock Dissipation and inter-scale transfer in fully coupled particle and
  fluid motions in homogeneous isotropic forced turbulence.
\newblock {\em International Journal of Heat and Fluid Flow}, 67:74--85.

\bibitem[Mehrabadi et~al., 2018]{Mehrabadi2018}
Mehrabadi, M., Horwitz, J., Subramaniam, S., and Mani, A. (2018).
\newblock A direct comparison of particle-resolved and point-particle methods
  in decaying turbulence.
\newblock {\em Journal of Fluid Mechanics}, 850:336--369.

\bibitem[Olbrich et~al., 2021]{Olbrich2021}
Olbrich, M., Bär, M., Oberleithner, K., and Schmelter, S. (2021).
\newblock Statistical characterization of horizontal slug flow using snapshot
  proper orthogonal decomposition.
\newblock {\em International Journal of Multiphase Flow}, 134.

\bibitem[Oseen, 1910]{Oseen1910}
Oseen, C. (1910).
\newblock Über die stokes'sche formel, und über eine verwandte aufgabe in der
  hydrodynamik.
\newblock {\em Arkiv för matematik, astronomi och fysik}, 6--29:1.

\bibitem[Pang and Wei, 2013]{Pang2013}
Pang, M. and Wei, J. (2013).
\newblock Experimental investigation on the turbulence channel flow laden with
  small bubbles by piv.
\newblock {\em Chemical Engineering Science}, 94:302--315.

\bibitem[Schiller and Naumann, 1933]{Schiller1933}
Schiller, L. and Naumann, A. (1933).
\newblock Über die grundlegenden berechnungen bei der schwerkraftaufbereitung.
\newblock {\em Zeitschrift des Vereines Deutscher Ingenieure}, 77:318--320.

\bibitem[Schneiders et~al., 2017]{Schneiders2017}
Schneiders, L., Meinke, M., and Schröder, W. (2017).
\newblock Direct particle-fluid simulation of kolmogorov-length-scale size
  particles in decaying isotropic turbulence.
\newblock {\em Journal of Fluid Mechanics}, 819:188--227.

\bibitem[Shen and Lu, 2021]{Shen2021}
Shen, J. and Lu, Z. (2021).
\newblock Influence of particle-fluid density ratio on the dynamics of
  finite-size particles in homogeneous isotropic turbulent flows.
\newblock {\em Physical Review E}, 104.

\bibitem[Sirovich, 1987]{Sirovich1987}
Sirovich, L. (1987).
\newblock Turbulence and the dynamics of coherent structures.
\newblock {\em Quarterly of Applied Math}, XLV(3):561--582.

\bibitem[Squires and Eaton, 1990]{Squires1990}
Squires, K. and Eaton, J. (1990).
\newblock Particle response and turbulence modification in isotropic
  turbulence.
\newblock {\em Physics of Fluids A: Fluid Dynamics}, 2:1191--1203.

\bibitem[Towne et~al., 2018]{Towne2018}
Towne, A., Schmidt, O., and Colonius, T. (2018).
\newblock Spectral proper orthogonal decomposition and its relationship to
  dynamic mode decomposition and resolvent analysis.
\newblock {\em Journal of Fluid Mechanics}, 847:821--867.

\bibitem[Viggiano et~al., 2018]{Viggiano2018}
Viggiano, B., Skjæraasen, O., Schümann, H., Tutkun, M., and Cal, R. (2018).
\newblock Characterization of flow dynamics and reduced-order description of
  experimental two-phase pipe flow.
\newblock {\em International Journal of Multiphase Flow}, 105:91--101.

\bibitem[Yeung and Pope, 1988]{Yeung1988}
Yeung, P. and Pope, S. (1988).
\newblock An algorithm for tracking fluid particles in numerical simulations of
  homogeneous turbulence.
\newblock {\em Journal of Computational Physics}, 79:373--416.

\end{thebibliography}
